# Substrate dopant induced electronic inhomogeneity in epitaxial bilayer graphene

*Shuai Zhang,[†] Di Huang,[†] Lehua Gu,[†] Yuan Wang,[†] Shiwei Wu[†\*]*

[†] State Key Laboratory of Surface Physics, Key Laboratory of Micro and Nano Photonic Structures (MOE), and Department of Physics, Fudan University, Shanghai 200433, China

**Abstract**

Two-dimensional (2D) materials have become a new territory for exploring novel properties and functionalities because of their superior tunability. The unprecedented tunability of 2D materials is also accompanied by many and equally great challenges, as they are susceptible to defects and disorders. The presence of defects and disorders induces the inhomogeneity of electronic states, often making it difficult to directly probe the intrinsic properties of materials. Therefore, many efforts have been devoted to improve the electronic homogeneity, for example, by reducing the density of defects and disorders in the materials and at the interface. However, little attention is paid to the disorders embedded in underlying substrates. Here we investigate how individual dopants in substrate interact with 2D materials and give rise to the electron-hole puddles by low temperature scanning tunneling microscopy (STM) and spectroscopy (STS).

Specifically, we take the epitaxial bilayer graphene grown on nitrogen doped silicon carbide (SiC) substrates as a model system, which has played the essential roles in many applications and fundamental studies. The nitrogen dopants inside SiC substrate were imaged over the epitaxial graphene by STM for the first time. The charged nitrogen dopants from the substrate induce the electron-lack puddles in graphene with a diameter of ~2 nm, via Coulomb interaction. The Fermi level with respect to the Dirac point is modulated by the puddles with an amplitude of ~40 meV, causing the electronic inhomogeneity of epitaxial graphene. Our findings on this prototype material are expected to facilitate the quality improvement of graphene and other 2D materials in general.

**Introduction**

The properties and functionalities of two-dimensional (2D) materials and devices heavily rely on the quality of the materials and their van der Waals interfaces.[1, 2] To improve the quality, quite a lot of efforts have been devoted. One major direction is to identify, understand, and control the type of defects in 2D materials,[3-5] with the ultimate goal to have the single crystalline, defect-free materials.[6-8] Another popular route is to obtain the clean interface by using ultraflat substrates such as hBN,[9-11] or reducing the corrugations including bubbles, ripples and wrinkles.[12, 13] However, the disorders embedded in the underlying substrates attract little attention.

In this work, we focus on the question how single dopants from the substrate affect the electronic inhomogeneity of 2D materials. To this end, we studied the epitaxial bilayer graphene grown on nitrogen doped silicon carbide (SiC) substrates by using low temperature scanning tunneling microscopy (STM) and spectroscopy (STS). Using a high bias voltage, we were able to visualize individual nitrogen dopants inside the SiC substrate. At low bias voltage, we characterized the electronic inhomogeneity of epitaxial graphene by measuring the tunneling spectroscopy right on or away from the substrate dopants. We found that the nitrogen dopants in SiC substrate cause the local electron-lack puddles in graphene via Coulomb interaction. Compared to the dopant-free region, the Fermi level with respect to the Dirac point is down shifted by about 40 meV. The size of electron-lack puddles in graphene is about 2 nm in diameter. We remarked that the fluctuation of charge carriers in hBN encapsulated graphene was reported to be $\sim 7 \times 10^{10}\ cm^{-2}$,[9] corresponding to tens of meV near the Dirac point. To further reduce the carrier fluctuation[14] and decrease the minimum conductance,[15] our finding infers that the defects inside the hBN substrate[16] must be taken into the proper consideration. Furthermore, the epitaxial bilayer graphene on doped SiC has been widely used as the substrate for molecular beam epitaxy (MBE) growth of other 2D materials such as transition metal dichalcogenides[17] and topological materials.[18] Our work suggests that the dopants in SiC may perturb the intrinsic properties of MBE-grown 2D materials.

**Result and Discussion**

**Sample treatment**

The growth of bilayer graphene on SiC substrate was carried out in a MBE system with a base pressure of $1 \times 10^{-10}$ torr. Commercially available nitrogen-doped 6H-SiC(0001) wafers with the resistivity of 0.02 ~ 0.1 $\Omega \cdot$ cm and the dopant density of about ~1x10$^{18}$ $cm^{-3}$ were used as substrate. After ultrasonic cleaning in acetone, the substrate was treated in ultrahigh vacuum by direct current heating. The growth of epitaxial bilayer graphene on Si-terminated 6H-SiC(0001) surface includes the following three sequential processes.[19-21] Firstly, the sample was degassed under 600 °C for 3 hours. Then, the sample was annealed at 800 °C under Si flux, which was supplied by direct current heating of a heavily doped silicon wafer. Thirdly, the sample was flash annealed without Si flux for several cycles. For every cycle, the sample was heated up to 1400 °C and held at high temperature for 5 minutes. The sample temperature was in situ monitored by an infrared pyrometer. During the annealing process, the chamber pressure was always maintained below $1 \times 10^{-9}$ torr. After the growth of bilayer graphene on SiC substrate, the sample was transferred from the MBE system to our STM system by a mobile suitcase with a base pressure of $3 \times 10^{-9}$ torr. Before the STM/STS measurements, the sample was outgassed again at 350 °C for 4 hours.

**Imaging substrate dopants**

Figure 1(a) shows the STM image of bilayer graphene epitaxially grown on 6H-SiC(0001) substrate at bias voltage of 4 V. As indicated by the parallelogram in Fig. 1(a), the (6×6) corrugation was clearly observed, consistent with previous STM studies.[22] Besides, pit-like new features emerged at this large bias voltage. When the same area was imaged at a lower bias voltage [Fig. 1(b)], these pit-like features disappeared. For comparison, Figure 1(c) plots the apparent height profiles across a same line in Fig. 1(a) and Fig. 1(b) at different sample bias voltages. It is seen that the depth of pits is larger than 150 pm at bias voltage of 4 V, while they are absent at bias voltage of 1 V.

To elucidate the origin of these pits, we measured atomically resolved STM topographic images. Figure 1(d) shows a STM topographic image of bilayer graphene at bias voltage of -0.8 V. On this image, two apparent periodicities are observed. The small period of 2.46 Å reflects the honeycomb lattice of graphene, marked by the hexagons in Fig. 1(d). The large period of 18.5 Å corresponds to the (6×6) corrugation observed in Fig. 1(a). Again, when the same area was imaged at bias voltage of 4 V, an apparent pit was observed [Fig. 1(e)]. We note that these pit-like features are distinct from the previous reported defects such as boundary loop[23] and point defect,[24, 25] which are embedded inside the bilayer graphene. In our sample, we also observed such graphene defects as, shown in Fig. 2. By comparison, these pit-like features do not originate from the defects or corrugation of bilayer graphene. Rather, they are likely from the defects underneath the bilayer graphene, i.e. inside the SiC substrate.

Since the SiC substrate is nitrogen doped, these pit-like features may come from the nitrogen dopants in SiC. According to electron spin resonance studies, the nitrogen dopants substitute the carbon sites in SiC,[26] as shown in Fig. 1(f). These nitrogen dopants can substitute at different layers with various subsurface depths. Indeed, the pit-like features in Fig. 1(a) exhibit different depth in STM topographic image. A more direct proof that the pits are from the nitrogen dopants would be the correlation between the density of pits in STM images and that of dopants in SiC substrate. According to the specification of SiC wafer, the bulk density of nitrogen dopants in SiC is about $\sim 10^{18}\ cm^{-3}$. If the nitrogen dopants within the subsurface depth of ~1 nm can be observed, the corresponding surface density of nitrogen dopants would be $10^{11} \sim 10^{12}\ cm^{-2}$. Experimentally, we estimated the pit density to be $[10^{11}-10^{12}]\ cm^{-2}$ from STM images, which does not depend on how the epitaxial bilayer graphene was grown. For example, Figure 3 shows the STM topography images of two bilayer graphene samples obtained under different growth conditions. They exhibited the same type of pit-like features and had identical density of pits regardless of the annealing time and rate. The consistency between the density of pits and the density of dopants permitted us to conclude that the observed pit-like features originate from the nitrogen dopants in SiC, which are underneath the epitaxial bilayer graphene.

**Electronic inhomogeneity of bilayer graphene**

To investigate the effect of nitrogen dopants on the electronic states, the dI/dV spectroscopic measurements were conducted on and off the dopants. Figure 4(a) shows the representative dI/dV spectra with a large range of sample bias voltage $V_b$. The dI/dV spectra exhibit almost vanishing conductance near zero bias voltage, which reflects the semiconducting nature of SiC underneath the graphene.[27] Around the sample bias voltage of 4 V, a peak appears on both dI/dV spectra. The energy of the peak on the dopant is higher than that obtained away from the dopant, consistent with the electron doping by nitrogen substitution to be discussed below. When the bias voltage was set to 4 V, the density of states on the dopant is smaller than that away from the dopant. Thus, the reduced tunneling conductance gives rise to the lower apparent height on the dopant in the STM topographic images shown in Fig. 1(a) and (e).

A straightforward question would be how the nitrogen dopants in SiC substrate affect the electronic property of epitaxial bilayer graphene. So we measured the dI/dV spectra with a small range of bias voltage $V_b$ and correlated the dI/dV spectra on and off the dopants, which are shown in Fig. 4(b). Overall, the dI/dV spectra are identical to previous reports.[20, 28, 29] Both the dI/dV spectra exhibit a dip around zero bias voltage, which could be attributed to the phonon-mediated tunneling.[30, 31] There is also a local minimum at negative bias voltage, corresponding to the Dirac point in graphene.[20, 29] From the dI/dV spectrum taken away from the dopants, the Dirac point is at -420 meV. By comparison, the Dirac point on the nitrogen dopant is shifted to -380 meV. This result suggests that the epitaxial graphene on SiC substrate is electron doped, and the nitrogen dopants reduce the level of electron doping in graphene. The shift of the Dirac point towards the Fermi level at the site of nitrogen dopants causes the larger density of state at bias voltage smaller than -420 meV (Fig. 4(b)). This effect is also seen by correlating the STM topography in Fig. 4(c) and the dI/dV spectroscopic image in Fig.

4(d). The nitrogen dopants appear as individual pits in STM topographic image, but they become bright protrusions in dI/dV spectroscopic image.

To understand the observed features, we sketched the scenarios with and without nitrogen dopants in Fig. 5(a). Since the nitrogen dopant substitutes the carbon atom in SiC, the nitrogen atom has one valance electron left after bonding with the neighboring silicon atoms. Then a net negative charge is localized around the dopant, causing electron or n-type doping in SiC. From the view of semiconductor energy band diagram, the negative charge of nitrogen dopants leads to upwards band bending,[32, 33] as shown in the upper panel in Fig. 5(a). This band bending of SiC is indeed observed in dI/dV spectra in Fig. 4(a). The peak of the local density of states on the nitrogen dopants shifts to higher sample bias voltage with respect to that on the defect-free region, by an amount of ~310 meV.

The Coulomb potential from the charged nitrogen substitution site tailors the local doping of graphene. For electrons above the Dirac point, the Coulomb interaction with the negatively charged nitrogen is repulsive. So the electron doping level in graphene near the nitrogen dopant is reduced, and the Fermi level shifts down towards the Dirac point,[34, 35] as shown in Fig. 5(b). Figure 5(c) illustrates the relationship between the sample bias voltage ($V_b$) and the local density of states (dI/dV) for both scenarios. At the nitrogen substitution site, the electron doping is lower, and the local density of states at the bias voltage slightly below the Dirac point ($V_D$) is higher. Thus, the nitrogen dopant in SiC induces inhomogeneous electron density landscape in graphene, effectively creating electron-lack puddles.[14]

**Conclusion**

In summary, we observed the nitrogen dopants in SiC substrate and unveiled their influence on the electronic inhomogeneity of epitaxial bilayer graphene by STM and STS. The negative charged nitrogen substitutes cause the shift of Dirac point in their vicinities and generate electron-lack puddles in graphene. Our work thus pinpoints the origin of electronic inhomogeneity in epitaxial bilayer graphene grown on SiC, paving the way to improve the quality of epitaxial graphene. This finding could also be extended to other 2D materials, where the electronic homogeneity is greatly affected by defects or dopants in the underlying substrate. Our result illustrates the use of nearly defect-free and ultraflat substrate can further improve the intrinsic properties of 2D materials.

**Methods**

**STM/STS measurements**

STM/STS measurements were performed in our home-built cryogen-free low STM system[36] at temperature of 16 K and a base pressure of $3 \times 10^{-11}$ torr. The sharp tungsten tip was fabricated by electrochemical etching, followed by self-sputtering in neon gas environment and annealing by electron beam bombardment. The tip was further conditioned and checked on a clean Au(111) surface before measurements.

While the tunneling current was collected on the tip, the bias voltage was applied to the sample. All the STM images were obtained in constant current mode. *dI/dV* spectra and mapping were measured by applying standard lock-in technique in open loop condition of the tunneling junction, while a sinusoidal modulation at frequency of 440 Hz with amplitude of 10 mV$_{rms}$ was added on the bias voltage.

**Corresponding Author**

* Email: swwu@fudan.edu.cn

**Author contributions**

S.W. conceived and supervised the project. S.Z. and Y.W. grew the samples. S.Z., D.H. and L.G. performed the measurements. S.Z. and S.W. analyzed the data and wrote the paper with contributions from all authors.

**Notes**

The authors declare no competing financial interests.

**ACKNOWLEDGMENT**

The work at Fudan University was supported by the National Basic Research Program of China (Grant Nos. 2016YFA0301002, 2019YFA0308404), National Natural Science Foundation of China (Grant No. 11427902).

**FIGUES and CAPTIONS**

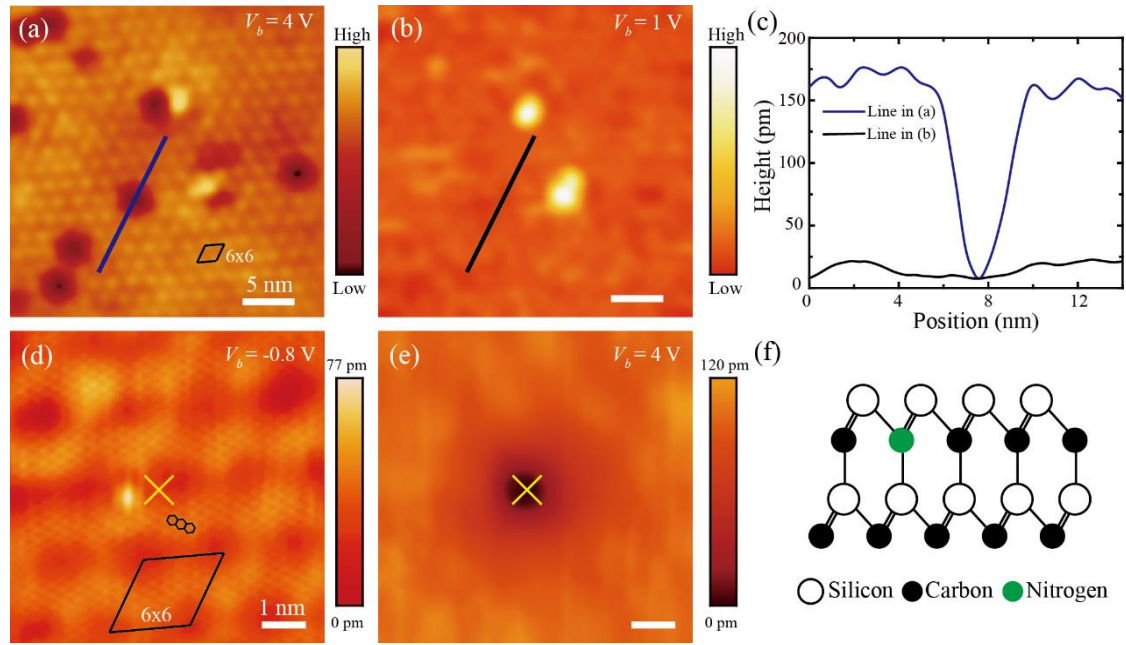

**Figure 1.** (a), (b) STM topography of epitaxial bilayer graphene on 6H-SiC(0001) at the same area, but under different bias voltage ($V_b$ = 4 V for a and $V_b$ = 1 V for b, respectively). $I_{set}$ = 100 pA. (c) Apparent height profiles along the line cut in a and b, respectively. (d) Atomically resolved STM image of the epitaxial bilayer graphene on 6H-SiC(0001) showing a (6×6) corrugation ($V_b$ = -0.8 V, $I_{set}$ = 100 pA). (e) STM image of the same region as d, but taken at $V_b$ = 4 V, $I_{set}$ = 100 pA. The yellow crosses in d and e indicate the position of a dopant. (f) Atomic structure of nitrogen doped SiC.

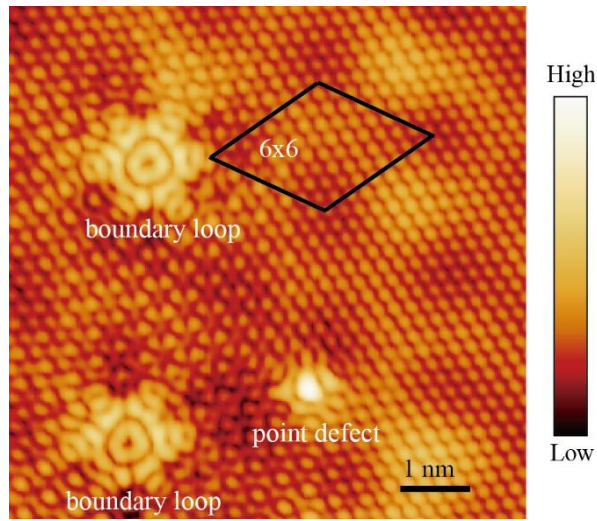

**Figure 2.** Atomic resolved STM image of bilayer graphene on 6H-SiC(0001). The image was taken at $V_b$ = -0.4 V, $I_{set}$ = 100 pA. The defects inside bilayer graphene including boundary loop and point defect were observed and marked.

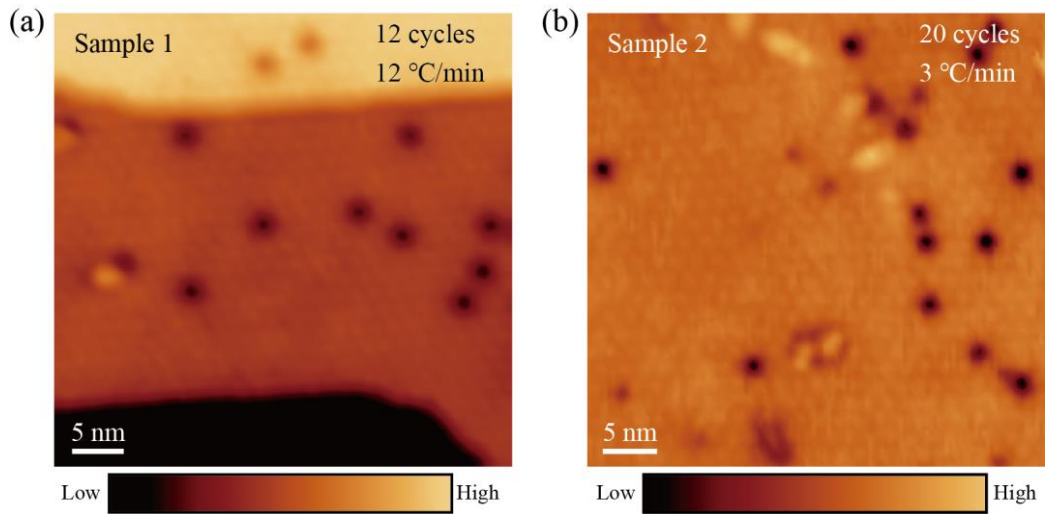

**Figure 3.** STM topography of epitaxial bilayer graphene on SiC under different conditions: (a) 12 cycles of flash annealing, cooling rate of 12 °C/min; (b) 20 cycles of flash annealing, cooling rate of 3 °C/min. The images were taken at $V_b$ = 4 V, $I_{set}$ = 100 pA. The dopants as pits exist in all the conditions.

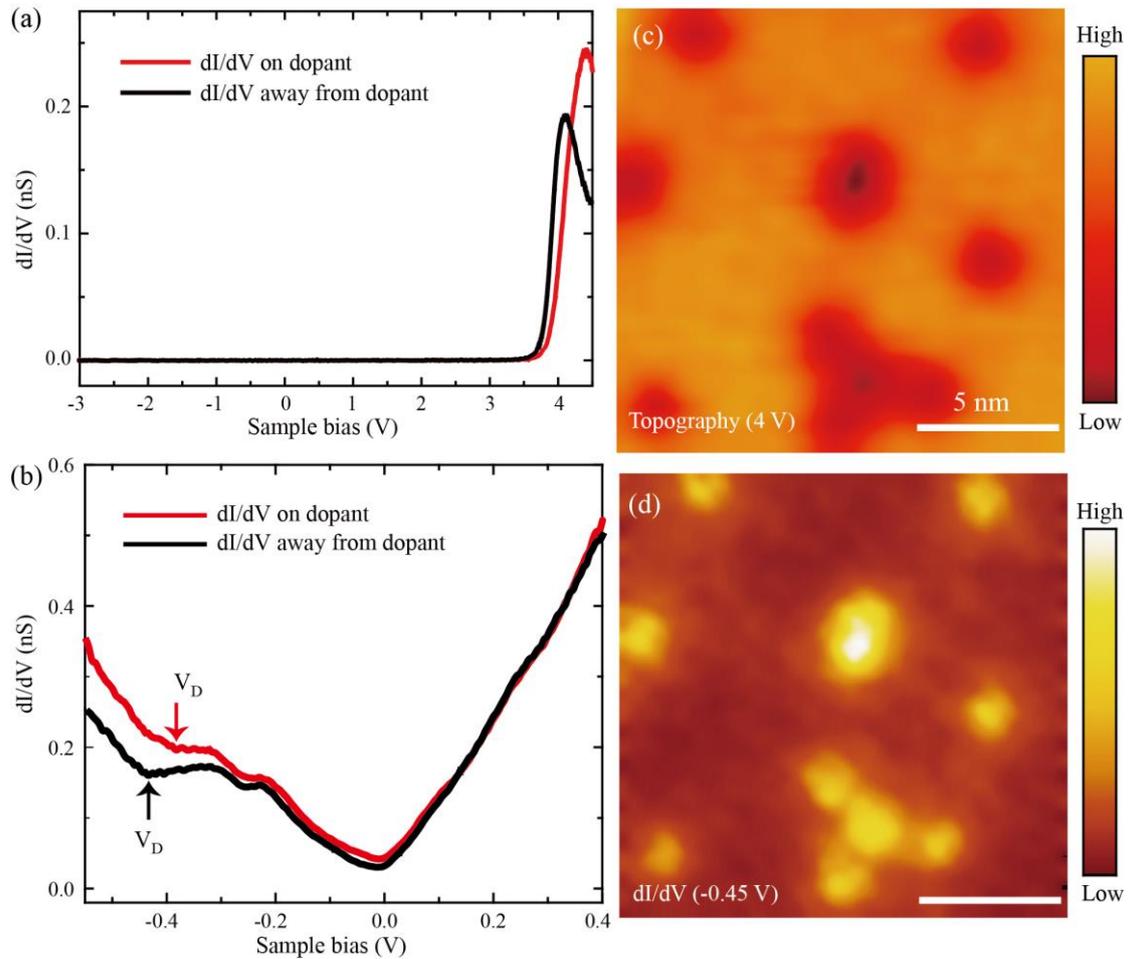

**Figure 4.** (a) Representative *dI/dV* spectra taken on and away from the individual dopants. The tunneling gap was set at $V_b$ = 4.5 V and $I_{set}$ = 100 pA. (b) Representative *dI/dV* spectra taken on and off a dopant in the bias voltage range from -0.55 V to 0.4 V. The tunneling gap was set at $V_b$ = 0.4 V and $I_{set}$ = 100 pA. The corresponding Dirac points were marked by black and red arrows. (c) STM image showing a region with several dopants ($V_b$ = 4 V, $I_{set}$ = 100 pA). (d) *dI/dV* ($V_b$ = -0.45 V) mapping acquired at the same region in c. The bright spots in d correspond to the pits or location of dopants in c.

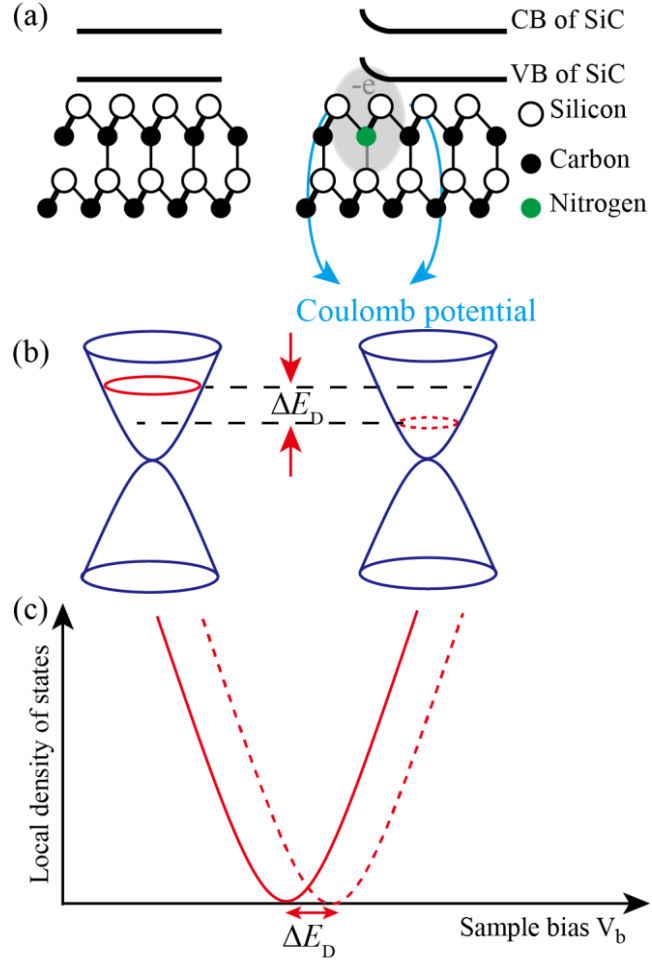

**Figure 5.** (a) Atomic structure and corresponding band diagram of SiC without and with nitrogen dopants, respectively. The nitrogen dopant causes electron doping of SiC and upward band bending. (b) Doping level of bilayer graphene on SiC. The negative charge of nitrogen dopant decreases the doping level of graphene due to the Coulomb potential. (c) Local density of states in graphene with the shifted Dirac point.